\documentclass[a4paper]{jpconf}
\usepackage{graphicx}
\usepackage{iopams}  
\begin{document}
\title{Blast wave fits with resonances to $p_t$ spectra from nuclear 
collisions at the LHC}

\author{Ivan Melo$^{a, b}$ and Boris Tom\'a\v{s}ik$^{b, c}$}

\address{$^a$ \v{Z}ilinsk\'a univerzita, Univerzitn\'a 1, 01026 \v{Z}ilina, 
Slovakia}
\address{$^b$ Univerzita Mateja Bela, FPV,  Tajovsk\'eho 40, 
97401 Bansk\'a Bystrica, Slovakia}
\address{$^c$ FNSPE, Czech Technical University in Prague, B\v{r}ehov\'a 7, 
11519 Prague 1, Czech Republic}

\ead{boris.tomasik@cern.ch}

\begin{abstract}
We report our results for the freeze-out temperature and transverse flow profile
obtained from fits to hadronic spectra measured by the ALICE collaboration. The 
influence of resonance decays is important and cannot be simply accounted for 
without the inclusion of their decays into the fits.
\end{abstract}

%%%%%%%%%%%%%%%%%%%%%%%%%%%%%%%%%%%%%%%%%%%%%%%%%%%%%%%%%%%%%%%%%%%%%%%%%

\section{Introduction}

Hadron spectra produced in ultrarelativistic nuclear collisions are shaped
by the dynamical state of the fireball at the moment of its breakup. 
The azimuthally integrated spectra are particularly characterised by the 
temperature and the transverse expansion velocity. 

We have analysed data by ALICE collaboration on transverse momentum 
spectra of $\pi$, $K$, $p$ \cite{spec_piKp}, $K_0$ and $\Lambda$ 
\cite{spec_K0Lam}, $\Xi$ and $\Omega$ \cite{spec_XiOm}, 
$K^*$ and $\Phi$ \cite{spec_KstarPhi}. For a single species
there is an ambiguity for the slope of the spectrum which  goes like 
\begin{equation}
\label{e:tstar}
T^* = T + m\langle v_t \rangle^2\,  .
\end{equation}
This is resolved by common fitting of different species.

Many hadrons originate from decays 
of resonances. They crucially influence the shape of the spectrum. We will 
show below that this cannot be accounted for  simply by shifting the 
extracted temperature. Also, it cannot be eliminated by choosing a 
fiducial $p_t$ interval for data fitting. 

The notorious problem of calculations with resonances is that they cannot 
be done reasonably analytically and one has to Monte-Carlo-simulate spectra 
which are then compared with real data. This is costly in terms 
of both CPU time and disk capacity. We have used the Monde Carlo event 
generator DRAGON \cite{dragon} for the generation of theoretical spectra.
It is a MC realisation of the blast wave model which was
adapted to be identical to the one used by ALICE collaboration for data 
fiting \cite{spec_piKp}.

%%%%%%%%%%%%%%%%%%%%%%%%%%%%%%%%%%%%%%%%%%%%%%%%%%%%%%%%%%%
\section{The model}

The blast wave model used in our fitting is characterised by the theoretical 
prescription 
\begin{equation}
\frac{dN}{dy\, d^2 p_t} = \int d\Sigma_\mu p^\mu \, f(p_\mu u^\mu,x) = 
\int d^4x\, S(x,p)\,  ,
\end{equation}
where $d\Sigma_\mu p^\mu$ indicate integration of the flux of particles 
through the three-dimensional freeze-out hypersurface and $f(p_\mu u^\mu , x)$ 
is the statistical distribution. The emission is conveniently expressed
by the emission function
\begin{equation}
\label{e:S}
S(x,p)\, d^4x = g_i\, \delta(\tau - \tau_{fo})\, m_t \cosh(y - \eta_s)
\Theta(R-r) \frac{1}{\exp\left (\frac{p_\mu u^\mu(x) - \mu(x)}{T} \right )\pm 1}\,
\tau d\tau \, d\eta_s\, r\, dr\, d\varphi\,  ,
\end{equation}
where we use $\tau = \sqrt{t^2-z^2}$, 
$\eta_s = \frac{1}{2}\ln \frac{t+z}{t-z}$,  polar coordinates
$r$, $\varphi$, and $g_i$ is spin degeneracy. The source is characterised 
by freeze-out Bjorken time $\tau_{fo}$ and transverse radius $R$. Statistical 
distributions are taken with the argument $p_\mu u^\mu$ which 
gives the energy in the local rest frame of the fluid. Here, $u_\mu$
is the local expansion four-velocity with longitudinal 
and transverse expansion
$
u^\mu = \gamma (1,\vec v) = (\gamma, \gamma v_t \cos\varphi,\gamma v_t \sin\varphi,
\gamma v_l)\,  ,
$
where $v_l = \tanh\eta_s$ and the transverse velocity grows with $r$ like
$
v_t(r) = \eta_f \left (  {r}/{R} \right )^n
$
with model parameters $\eta_f$ and $n$. The mean transverse
velocity is
$
\langle v_t \rangle = 2{\eta_f}/{(1+n)}
$.

Resonances are also produced according to the emission function (\ref{e:S}).
Then they are let to decay with the assumption of constant matrix elements
and uniform filling of the phase space.

%%%%%%%%%%%%%%%%%%%%%%%%%%%%%%%%%%%%%%%%%%%%%%%%%%%%%%%%%%%%%%%%%%%%%%%%%
\section{The influence of resonances}

Chemical composition of the fireball plays 
a crucial role as it determines which part of the final state hadrons 
is due to resonance decays. We apply here the 
scenario of two freeze-outs: chemical freeze-out at a higher temperature 
is followed by further cooling and expansion which ends up in the thermal
freeze-out. The abundances of all species---including resonances---are 
determined by chemical equilibrium at chemical freeze-out temperature
of 152~MeV and baryochemical potential of 1~MeV \cite{chemfit}.
This assumption is rather extreme and actually means that the kinetic 
freeze-out follows the chemical so quickly that no resonances have enough 
time to decay. 

Often, resonance contributions are left out in the fits to transverse 
momentum spectra. It is argued then that 
this simplification is justified if the fit is performed in a fiducial region 
where $p_t$ is high enough so that resonance contributions are 
not important and still low enough so that hard processes play no role. 
Unfortunately, the common lore that resonance contribution is concentrated 
only at low $p_t$ is wrong! We have checked this by performing fits 
to fiducial $p_t$-intervals used in \cite{spec_piKp}.

Some of our results are summarised in Table \ref{t:compar}.
%
%%%%%%%%%%%%%%%%%%%%%%%%%%%%%%%%%%%%%%%%%%%%%%%%%%%%%%%%%%
\begin{table}[t]
\begin{center}
\caption{\label{t:compar}
Selected fit results using the fiducial intervals 
as determined by the ALICE collaboration 
\cite{spec_piKp}. Compared are sets of results with and without inclusion of resonance 
decays.}
\lineup
\begin{tabular}{lllllll}
\br
 &  \centre{3}{no resonances} & 
\centre{3}{with resonances} \\ \ns\ns
&  \crule{3} & \crule{3}\\
centrality & 
\parbox{2.7em}{$T$ (MeV)} & $\langle v_t \rangle$ & $n$ & 
\parbox{2.7em}{$T$ (MeV)} & $\langle v_t \rangle$ & $n$\\
\mr
\00--5\% & \098& 0.645& 0.73& \082& 0.662& 0.69\\
%\05--10\% & \098 & 0.645 & 0.73 & \094 & 0.654 & 0.69\\
10--20\% & 102 & 0.637 & 0.73 & \090 & 0.649 & 0.71 \\
%20--30\% & 102 & 0.624 & 0.79 & \098 & 0.633 & 0.75 \\
30--40\% & 110 & 0.605 & 0.81 & 102 & 0.616 & 0.79 \\
%40--50\% & 110 & 0.572 & 0.97 & 118 & 0.581 & 0.89 \\
50--60\% & 122 & 0.527 & 1.15 & 126 & 0.541 & 1.03\\
%60--70\% & 126 & 0.484 & 1.39 & 146 & 0.489 & 1.23 \\
70--80\% & 142 & 0.439 & 1.51 & 170 & 0.423 & 1.55 \\
\br
\end{tabular}
\end{center}
\end{table}
%%%%%%%%%%%%%%%%%%%%%%%%%%%%%%%%%%%%%%%%%%%%%%%%%%%%%%%%%%
%
Our fit results with no resonances included agree
with those by ALICE \cite{spec_piKp}. The inclusion of resonances, however, 
\emph{does} change the results. Moreover, the change depends on centrality.
While in central collisions the temperature \emph{drops} by 10~MeV when 
resonances are included, in peripheral collisions it \emph{rises} by 
10~MeV.

\begin{figure}[t]
\centerline{\includegraphics[width=0.75\textwidth]{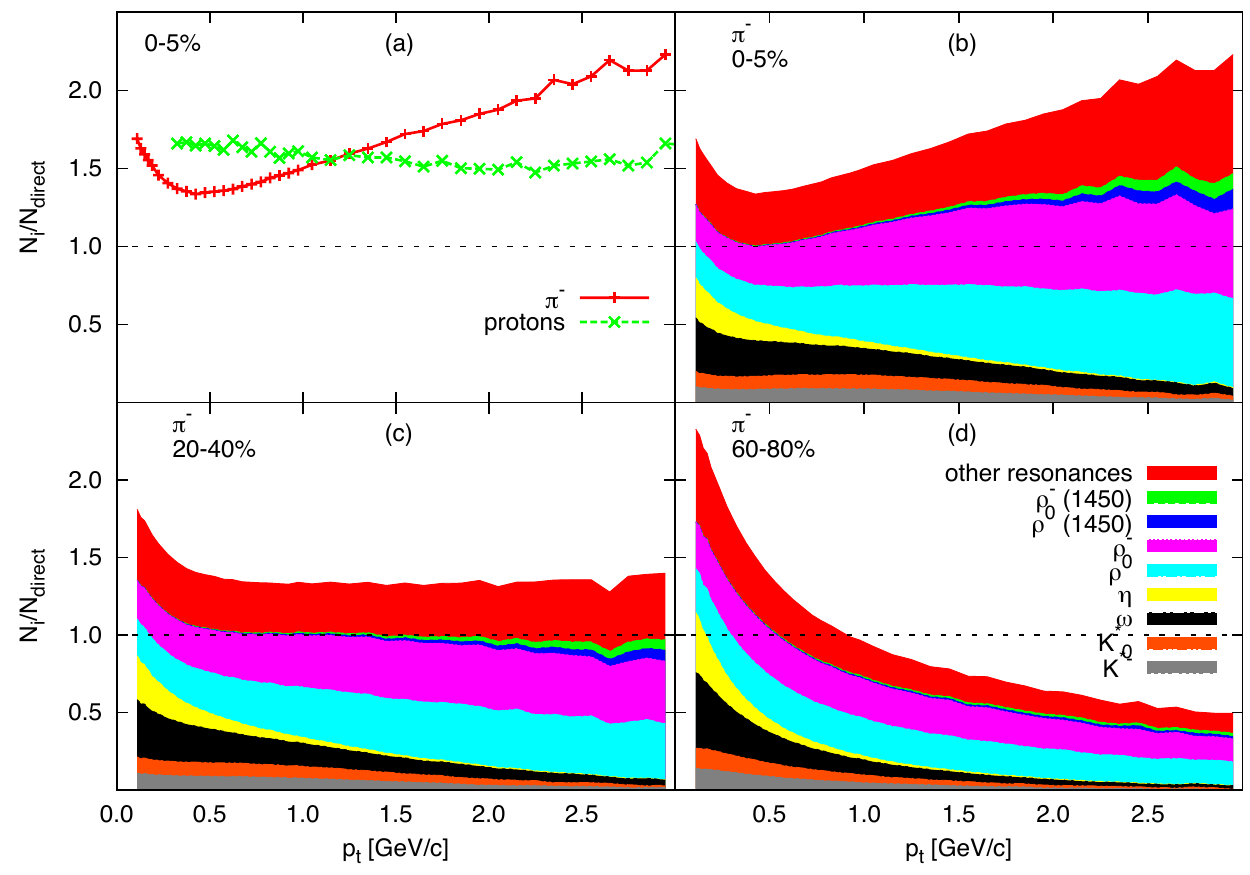}
}
\caption{\label{f:anat}
The ratio of hadrons from resonance decays to those produced directly as a function 
of $p_t$. (a) pions and protons in 0--5\% central collisions; (b)-(d) the ratio is broken 
up into contributions from individual resonance species, different panels show 
different centralities. Simulations correspond to our best fits to the indicated centralities.
The used values of $T$ and $\langle v_t\rangle$ are indicated in Fig.~\ref{f:results}.}
\end{figure}

To see the cause of such shifts, we show in Fig.~\ref{f:anat} how the 
resulting spectra are build up from contributions of individual resonance 
species.
%
%%%%%%%%%%%%%%%%%%%%%%%%%%%%%%%%%%%%%%%
%%%%%%%%%%%%%%%%%%%%%%%%%%%%%%%%%%%%%%%
%
This was calculated with our best fits to the given centrality 
classes, as reported below.
In peripheral collisons we indeed observe that resonance production 
populates mainly the low $p_t$ region. This feature still survives in the  
20--40\% centrality class. In the most central collisions, however,  
the relative contribution from resonance decays grows with increasing $p_t$. 

The figure also demostrates the reason for such a behaviour. 
In central collisons temperature is low and transverse flow strong. 
Low $p_t$ region is populated by decays of $\eta$ and $\omega$. The former 
is light and the latter decays into three pions. Hence, there is not 
much energy left for the momentum of the pions. The higher-$p_t$ tail 
is mainly populated by decays of $\rho$ and heavier resonances. When they 
decay, much energy is available for the kinetic energy of the pions.
Pions get a kick from decays of such resonances.  
Also, due to low temperature not many direct pions have high $p_t$ and so 
the share of  pions from heavy resonance decays grows. Strong transverse 
flow boosts also the heavy resonances and helps to pronounce the kick 
to high $p_t$. This changes when the temperature grows and the transverse 
expansion is weaker (peripheral collisions). There,  directly produced pions 
can better populate also higher $p_t$ and the share of resonance production 
at high $p_t$ does not grow anymore. 

Qualitatively similar results for resonance production, although not so 
much pronounced as for pions, are obtained for protons, as well. 
(Shown in the supplementary online material.)

Thus there is i) no fiducial interval where resonance decay contribution
can be ignored; ii) no simple recipe stating that the inclusion of resonances 
will just shift the fit results by any fixed value, since 
even qualitatively the influence of resonance decays depends on the state 
of the fireball. 

%%%%%%%%%%%%%%%%%%%%%%%%%%%%%%%%%%%%%%%%%%%%%%%%%%%%%%%%%%%%%%%%%%%%%%%%%%%
\section{Fit results}

\begin{figure}[t]
\includegraphics[width=0.5\textwidth]{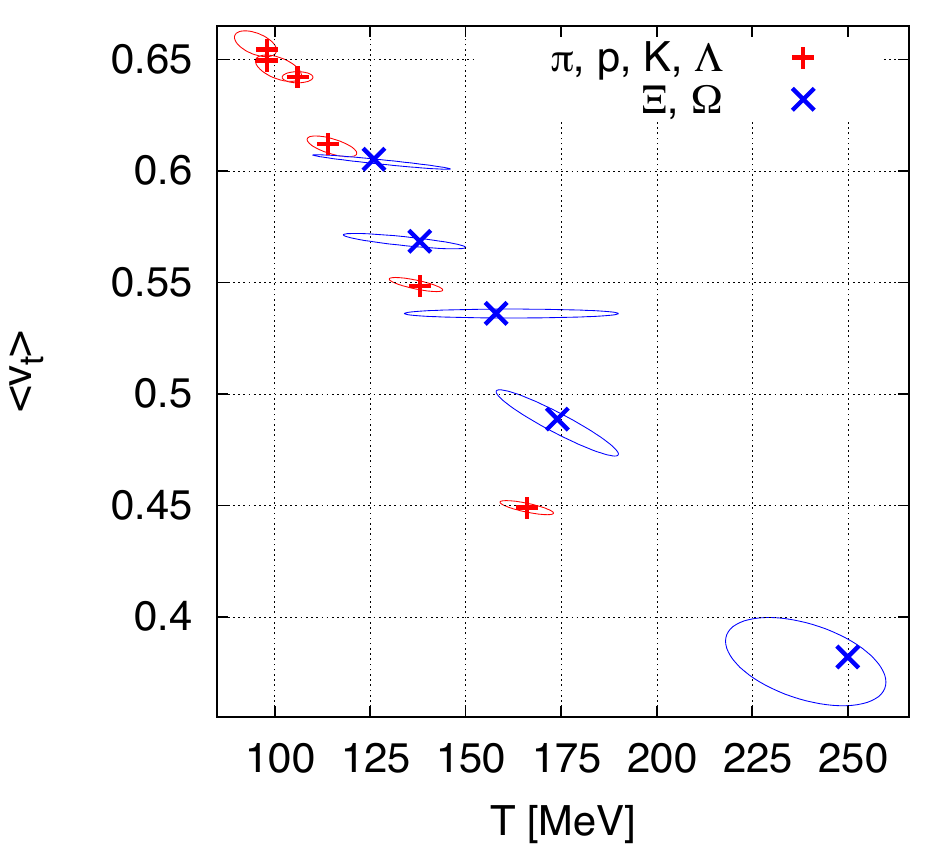}\hspace{0.05\textwidth}%
\begin{minipage}[b]{0.45\textwidth}\caption{\label{f:results}
Summary of the fitted temperatures and transverse flows.  
Different values show results for different 
centralities with top left being the most central and bottom right the most peripheral. 
The centralities for common fits to 
$\pi$, $K$, $p$, and $\Lambda$ spectra (symbol $+$) are: 0--5\%, 5--10\%, 10--20\%, 
20--40\%, 40--60\%, 60--80\%. Centralities for $\Xi$ and $\Omega$ spectra
(symbol $\times$) are:
0--10\%, 10--20\%, 20--40\%, 40--60\%, 60--80\%. Elipses indicate our estimates of 
99\% confidence levels (for  $\pi$, $K$, $p$, and $\Lambda$) and 1$\sigma$ intervals
(for $\Xi$ and $\Omega$). 
}
\end{minipage}
\end{figure}

The obtained freeze-out temperatures grow and the transverse expansion 
velocity decreases as we move from central collisions to more peripheral. 
The overview of results 
can be found in Figure~\ref{f:results}. 
%
%%%%%%%%%%%%%%%%%%%%%%%%%%%%%%%%%%%%
%%%%%%%%%%%%%%%%%%%%%%%%%%%%%%%%%%%%
%
With the model fitted to pions, kaons, nucleons, and $\Lambda$'s 
we were able to reproduce also the $K^*$ and $\phi$ spectra. However, 
the spectra of multistrange hadrons, $\Xi$ and $\Omega$ did not 
fall into this systematics. When we fitted them separately, the obtained 
temperatures were generally higher and transverse flows generally weaker 
than those of more abundant species. 

More detailed description of the procedure and results can be found in 
\cite{fitpaper}.

%%%%%%%%%%%%%%%%%%%%%%%%%%%%%%%%%%%%%%%%%%%%%%%%%%%%%%%%%%%%%%%%%%%%%%%%%%

\section{Conclusions}

Our results indicate that between the chemical and thermal freeze-out 
the temperature drops considerably in central collisions. It opens the question 
whether assuming abundances of all resonances given by chemical 
freeze-out temperature is justified. We shall reconsider it soon.

It should also be noted that a successful fit on $p_t$ spectra including 
resonances and assuming common chemical and thermal freeze-out was recently 
reported in \cite{begun}.

%%%%%%%%%%%%%%%%%%%%%%%%%%%%%%%%%%%%%%%%%%%%%%%%%%%%%%%%%%%%%%%%%%%%%%%%%%%

\subsection*{Acknowledgements}

Supported in parts by APVV-0050-11, VEGA 1/0469/15 (Slovakia) and 
M\v{S}MT grant LG13031 (Czech Republic). 
Computing performed in HPCC of Matej Bel University
supported by grants ITMS 26230120002 and 26210120002.
BT acknowledges the support of attendance to this conference by the Plenipotentiary 
of the Slovak Government to JINR Dubna.

%%%%%%%%%%%%%%%%%%%%%%%%%%%%%%%%%%%%%%%%%%%%%%%%%%%%%%%%%%%%%%%%%%%%%%%%%%%
%%%%%%%%%%%%%%%%%%%%%%%%%%%%%%%%%%%%%%%%%%%%%%%%%%%%%%%%%%%%%%%%%%%%%%%%%%%

\pagebreak

Supplementary material shows the ratios of hadrons from resonance decays 
to those produced directly as a function of $p_t$ for protons in three 
centrality classes. The plots show the composition from various kinds of resonances.

\centerline{\includegraphics[width=0.64\textwidth]{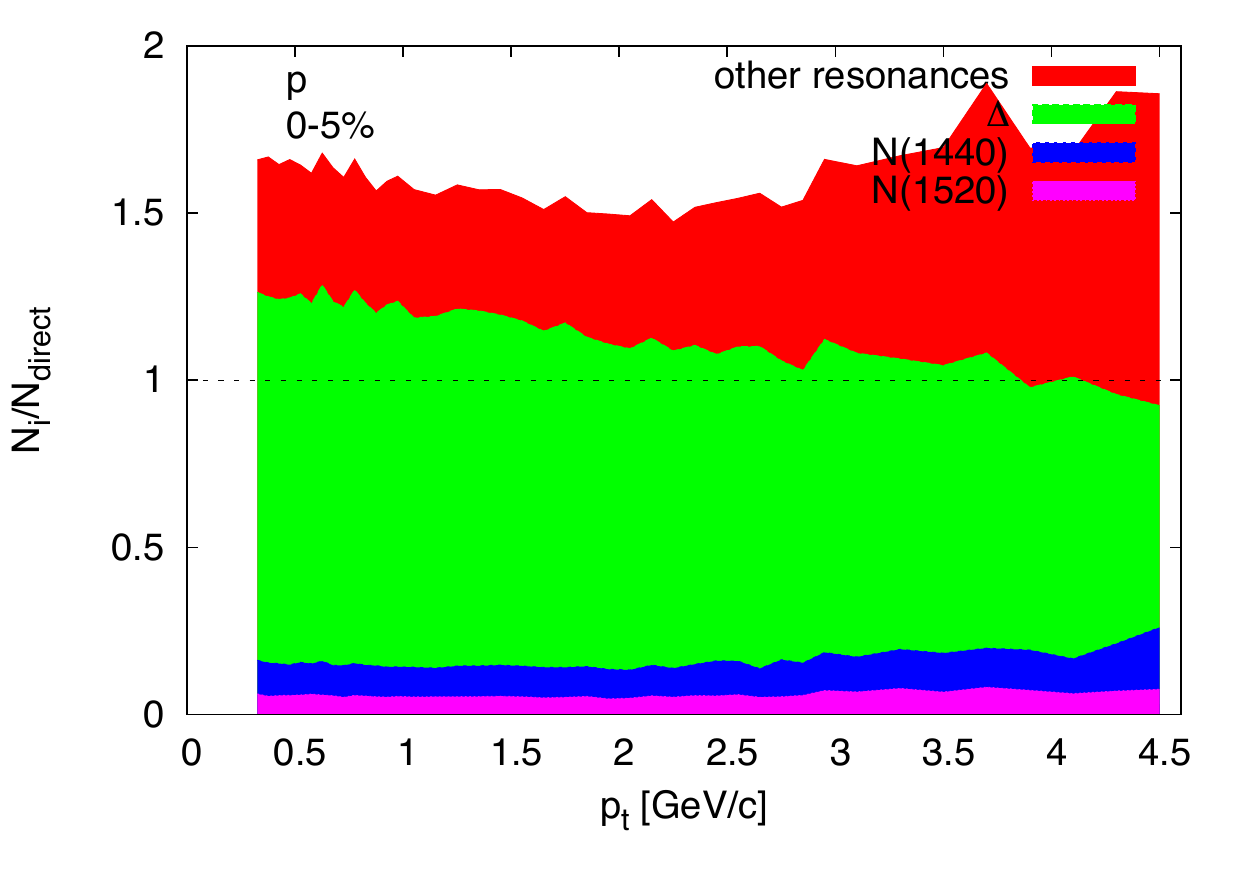}}
\centerline{\includegraphics[width=0.64\textwidth]{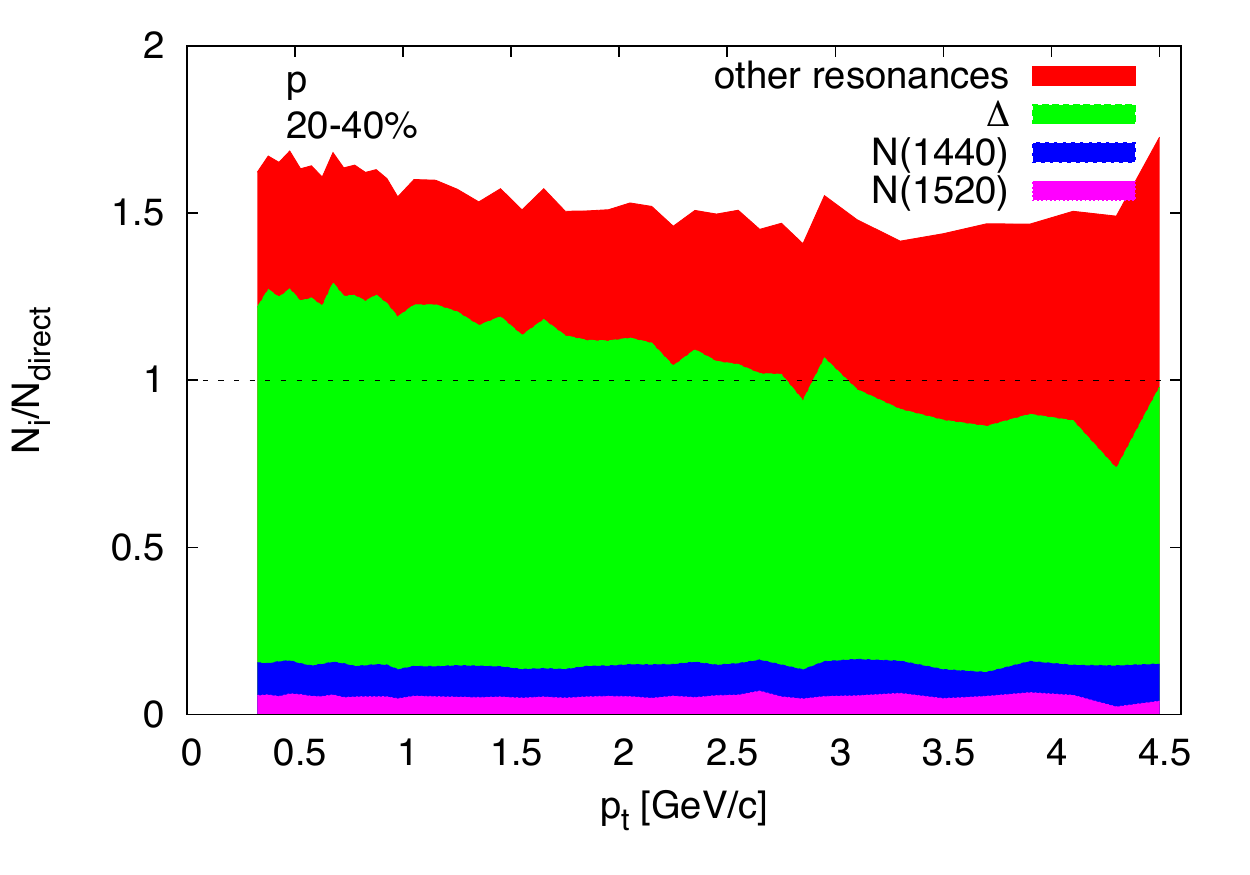}}
\centerline{\includegraphics[width=0.64\textwidth]{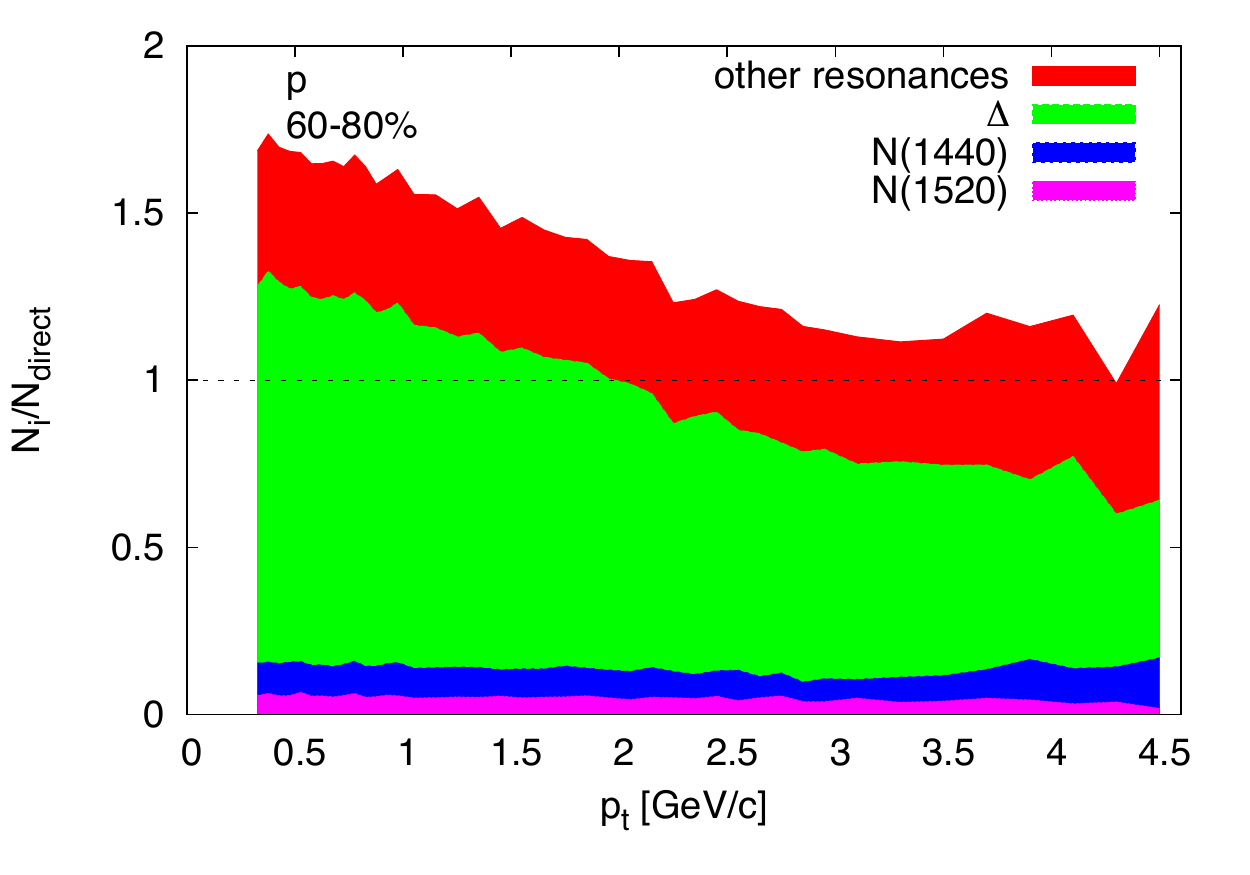}}

\end{document}